\documentclass[aps,prl,10pt,twocolumn,showpacs,superscriptaddress]{revtex4-1}
\usepackage{amsmath}
\usepackage{latexsym}
\usepackage{amssymb}
\usepackage{bm}
\usepackage{graphics,epstopdf}
\usepackage{color}
\usepackage{hyperref}

\usepackage{newlfont}
\usepackage{amsfonts}
\usepackage{amsthm}
\usepackage{graphicx}
\usepackage{epsfig}

\newcommand{\ket}[1]{|{#1}\rangle}

\newcommand{\expec}[1]{\langle#1\rangle}

\usepackage{times}

\usepackage[up]{subfigure}

\newcommand{\be}{\begin{equation}}
\newcommand{\ee}{\end{equation}}
\newcommand{\bc}{\begin{center}}
\newcommand{\ec}{\end{center}}
\newcommand{\bea}{\begin{eqnarray}}
\newcommand{\eea}{\end{eqnarray}}
\newcommand{\ba}{\begin{array}}
\newcommand{\ea}{\end{array}}

\begin{document}
\title{Polarization Squeezing in Degenerate Parametric Amplification of Coherent Light}

\author{Namrata Shukla}
\email{namratashukla@hri.res.in}

\affiliation{Now at, Quantum Information and Computation Group,\\
Harish-Chandra Research Institute, Chhatnag Road, Jhunsi,
Allahabad 211 019, India}

\affiliation{Department of Physics,\\
University of Allahabad, Allahabad, Allahabad-211001, UP, India}

\author{Ranjana Prakash}


\affiliation{Department of Physics,\\
University of Allahabad, Allahabad, Allahabad-211001, UP, India}



\date{\today}

\begin{abstract}
We study polarization squeezing of a light beam initially in the coherent state using the nonlinear interaction hamiltonian 
$ H=k\big(\hat a_{x}^{\dagger2}+{\hat a_{x}}^2\big)$. For the degree of polarization squeezing, 
we use a definition written in the final form by the authors and also used in earlier papers.
We find that the polarization squeezing can be very high and the degree of polarization squeezing 
can be less than unity by a very small amount. We achieve the polarization squeezing even for very 
low beam intensity under some conditions involving phase angles in the polarization modes.

\end{abstract}

\maketitle
\section*{Introduction}
Polarization of light was initially defined with the help of Stokes parameters \cite{1,2} $(S_1, S_2, S_3)$.
The point having coordinates $(S_1, S_2, S_3) $ lies on a sphere of radius $S_0$ called Poincare sphere
represented by
\begin{equation}
\label{eq1}
\bm S^2=S_{0}^2=S_{1}^2+S_{2}^2+S_{3}^2.
\end{equation}
For unpolarized light $ |{\bm S}|=0 $ and for partially polarized light $ S_0>|{\bm S}| $.
Stokes parameters involve coherence functions\cite{3} of order (1,1) and it 
has been realized that these are insufficient to describe polarization completely 
as $\bm{S}=0$ does not represent only unpolarized light.\cite{4}
These parameters are still relevant because of the non-classicalities associated with polarization, 
{\it viz}, the polarization squeezing and polarization entanglement.\cite{5} \\ 

Appropriate continuous variables for non-classical polarization states are Stokes 
operators\cite{6} which have a clear analogy with the Stokes parameters. These operators 
$\hat S_{0}$ and ${\hat{\bm S}}=\hat S_{1}, \hat S_{2}, \hat S_{3}$ have a similarity with spin variables of atomic systems 
and are given by
\begin{equation}
\label{eq2}
\hat S_{0, 1}=\hat a_{x}^\dagger \hat a_{x}\pm \hat a_{y}^\dagger \hat a_{y},~ \hat S_{2}+i \hat S_{3}=2 \hat a_{x}^\dagger \hat a_{y}.
\end {equation}
Simultaneous exact measurement of these variables is impossible as they follow the uncertainty 
relations to limit their variances written as
\begin{equation}
\label{eq3}
V_{j} V_{k}\geqslant{\expec{\hat S_{l}}}^2,~ 
V_{j}\equiv\expec{\hat S_{j}^2}- {\expec{\hat S_{j}}}^2,
\end{equation}
which can be obtained by using the commutation relations
\begin{equation}
\label{eq4}
[\hat S_0, \hat S_j]=0,~[\hat S_j, \hat S_k]=2i{\sum_{l}\epsilon_{jkl}}~\hat S_{l}.
\end {equation}
$\epsilon_{jkl}$ being Levi-Civita symbol for $(j,k,l=1,2,$ or $3,~j\neq k\neq l\neq j)$ as annihilation and creation operators follow the commutation relations
$ [\hat a_{j}, \hat a_{k}^\dagger]=\delta_{jk}$.\\

Similar to the concept of ordinary squeezing, polarization squeezing is defined using the 
commutation relations followed by Stokes operators and uncertainty products. 
Polarization squeezing having applications in quantum information theory, it is desirable to devise methods for generation of 
states with appreciable polarization squeezing. \\

In this paper, we investigate the polarization squeezing using the most general criterion for it, 
in the coherent light under a squeezing operation such as degenerate parametric amplification with a nonlinear interaction hamiltonian and report the closed form result. 
The incident coherent beam shows high polarization squeezing after nonlinear interaction and 
the degree of polarization squeezing can be less than unity by a very small amount for certain photon intensity and phases in two polarization modes. 
It is important to note that in most cases of realizations of polarization squeezing the beam intensity has been very large in order to efficiently realize a quadrature measurement. 
However, in this study we get a tighter bound on beam intensity with some additional conditions.

\section*{Criterion for Polarization squeezing}
Polarization squeezing was first defined by Chirkin et al.\cite{7} as
\begin {equation}
\label{eq5}
 V_{j}<V_{j} (coh)=\hat S_{0},
\end {equation}
,{\it i.e.}, $\hat S_{j}$ is squeezed if the variance of a Stokes operator is less than that for 
an equally intense coherent state.\\

Heersink et al.\cite{8} in their earlier mentioned paper, defined polarization squeezing using the uncertainty 
relations given by Eq.~(\ref{eq3}) in the form
\begin {equation}
\label{eq6}
 V_{j}<\mid\expec{\hat S_{l}}\mid<V_{k} ~~for~~ j\neq k \neq l,
\end {equation}
for squeezing of component $\hat S_{j}$ of Stokes operator vector. \\

Luis and Korolkova\cite{9} finally compared various criteria and gave 
the following criterion for polarization squeezing of a component of $\hat {\bm S}$ along 
a unit vector ${\bm n}$ as
\begin{equation}
\label{eq7}
 V_{\bm n}<\mid\expec{\hat S_{\bm n_{\perp}}}\mid,
\end{equation}
where $ \hat S_{\bm n_{\perp}}$ is component of $\hat {\bm S}$ along unit vector $\bm n_{\perp}$ 
which is perpendicular to $ \bm n $. For suitable orthogonal components $ \hat S_{\bm n} $ and 
$ \hat S_{\bm n_{\perp}}$, they have discussed the order of stringency of the various criteria as
\begin{equation}
\label{eq8}
 V_{\bm n}<\expec{\hat S_{\bm n_{\perp}}}^2 / \expec{\hat S_{0}}
 <\mid\expec{\hat S_{\bm n\perp}}\mid<\expec{\hat S_0}.
\end{equation}

The authors finally have written the criterion for polarization squeezing in 
the form \cite{10} 
\begin{eqnarray}
\label{eq9}
V_{\bm n}\equiv\expec{\Delta \hat S_{\bm n}^2} &< ~~~~ {|\expec{\hat S_{\bm n_{\perp}}}|}_{max} 
\nonumber \\
&~~~~ =\sqrt{{|\expec{\hat {\bm S}}|}^2-{\expec{\hat S_{\bm n}}}^2}.
\end{eqnarray}
arguing that there are infinite directions $ \bm n_{\perp} $ for a given component $ \hat S_{\bm n} $
and therefore it is needed to consider maximum possible value of $ \mid\expec{\hat S_{n_{\perp}}}\mid $.  
We shall use the criterion in Eq.~(\ref{eq9}) for polarization squeezing, which is most general and based 
on the actual uncertainty relations. We define polarization squeezing factor $ \mathcal{S}_{\bm n}$ and 
degree of polarization squeezing $ \mathcal{D}_{\bm n} $ as
\begin{equation}
\label{eq10}
\mathcal{S}_{\bm n} =\frac{V_{\bm n}}{\sqrt{{\mid\expec{\hat {\bm S}}\mid}^2-{\expec{\hat S_{\bm n}}}^2}},~
\mathcal{D}_{\bm n}=1-\mathcal{S}_{\bm n}.
\end{equation}
Non-classicalities are observed when $ 1>\mathcal{S}_{\bm n}>0 $ and the degree of polarization squeezing 
$ \mathcal{D}_{\bm n}$ lies between $0$ and $1$. 

\section*{Hamiltonian and Polarization squeezing}
Consider a beam of two mode coherent radiation and we pass it through a squeezing operation like 
degenerate parametric amplification that changes $x$ mode after the nonlinear interaction.
Hamiltonian in the interaction picture\cite{11} for 
this kind of a squeezing operation can be written as
\begin{equation}
\label{eq11} H=k\big(\hat a_{x}^{\dagger2}+{\hat a_{x}}^2\big),
\end{equation}
where $ \hat a_{x,y} $ are annihilation operators for the two linear polarizations $x$ and $y$. 
The state at time $t$ after this nonlinear interaction is given by
\begin{eqnarray}
\label{eq12}
 \hat a_x(t)&=&c~\hat a_x-is~\hat a_{x}^\dagger, \nonumber \\
 \hat a_y(t)&=&\hat a_y.
\end{eqnarray}
where $c=\cosh 2kt,~s=\sinh 2kt$, $k$ being a coupling constant.\\

If, the incident light in the coherent state $ \ket{\alpha, \beta}$ with
$ \alpha=A\cos{\theta}~e^{i\phi_x},~\beta=A\sin{\theta}~e^{i\phi_y}$, straight forward 
calculations give the expectation values and variances of Stokes operators 
after the interaction time $ kt $ as
\begin{eqnarray}
\label{eq13}
\expec{\hat S_1}&=&(c^2+s^2)|\alpha|^2-|\beta|^2-2cs\big(|\alpha|^2\sin2\phi_x\big),\nonumber\\
\expec{\hat S_2}&=&2c|\alpha||\beta|\cos(\phi_x-\phi_y)-2s|\alpha||\beta|\sin(\phi_x+\phi_y),\nonumber\\
\expec{\hat S_3}&=&2s|\alpha||\beta|\cos(\phi_x+\phi_y)-2c|\alpha||\beta|\sin(\phi_x-\phi_y),\nonumber\\
\end{eqnarray}
and 
\begin{eqnarray}
\label{eq14}
V_1&=&(c^2+s^2)|\alpha|^2+2c^2s^2\big(2|\alpha|^2+1\big)+|\beta|^2\nonumber\\ &&-4cs(c^2+s^2)|\alpha|^2\sin2\phi_x,\nonumber\\
V_2&=&c^2\big(|\alpha|^2+|\beta|^2\big)+s^2\big(|\alpha|^2+|\beta|^2+1\big)\nonumber\\ 
&&-2cs\big(|\alpha|^2\sin2\phi_x+|\beta|^2\sin2\phi_y\big),\nonumber\\
V_3&=&c^2\big(|\alpha|^2+|\beta|^2\big)+s^2\big(|\alpha|^2+|\beta|^2+1\big)\nonumber\\
&&-2cs\big(|\alpha|^2\sin2\phi_x+|\beta|^2\sin2\phi_y\big).\nonumber\\
\end{eqnarray}

In order to study the maximum polarization squeezing, we perform extensive numerics for the minimization of 
squeezing factors obtained by plugging in the values for all three Stokes operators $\hat S_1 $, $ \hat S_2 $ and $\hat S_3 $ using the criterion mentioned in Eq.~(\ref{eq10}). 
This minimization was done for fixed number of photons in both the polarization modes and we observe that with the forms of polarization squeezing factors obtained for set of values $(\phi_x,\phi_y)$ 
corresponding to maximum polarization squeezing in $\hat S_1 $ and $\hat S_3 $, it is challenging to perform analytical calculations. \\

However, we find that $\mathcal{S}_{2}$ attains its minimum at $\big(\phi_x,\phi_y)=\big(\frac{\pi}{4},\frac{\pi}{4}\big), 
\big(\frac{5\pi}{4},\frac{\pi}{4}\big), \big(\frac{\pi}{4},\frac{5\pi}{4}\big)$ and $\big(\frac{5\pi}{4},\frac{5\pi}{4}\big)$ making it analytically feasible to find the conditions 
on photon intensity and time for maximum polarization squeezing to occur, though it is difficult to show that these are the only minima. 
The polarization squeezing factor and degree of polarization squeezing for Stokes operator $\hat S_2$ has the form
\begin{equation}
\label{eq15}
\mathcal{S}_{2}=\frac{V_2}{\sqrt{{\expec{\hat S_1}}^2+{\expec{\hat S_3}}^2}}, ~\mathcal{D}_{2}=1-\mathcal{S}_{2}.
\end{equation}
For these set of values $(\phi_x,\phi_y)$, the expression for $\mathcal{S}_{2}$ can be written as
\begin{equation}
\label{eq16}
\mathcal{S}_{2}=\frac{(c-s)^2\big|(\alpha|^2+|\beta|^2)+s^2}{\big|(c-s)^2|\alpha|^2-|\beta|^2\big|}.
\end{equation}
or
\begin{equation}
\label{eq17}
\mathcal{S}_{2}=\frac{Y}{|X|},
\end{equation}
where $ X=\big|(c-s)^2|\alpha|^2-|\beta|^2\big|,~Y=(c-s)^2(\big|\alpha|^2+|\beta|^2)+s^2$.
Polarization squeezing, {\it i.e.}, $(\mathcal{S}_{2}<1)$ is not likely to occur when $X>0$ as $Y-|X|=(1+e^{-4kt})$ 
can not take negative values. For polarization squeezing, therefore, we look for $X<0$ and $|X|=-X$. Let us write
\begin{equation}
\label{eq18}
 X=|\alpha|^2(e^{-4kt})+\frac{1}{4}\big(e^{4kt}+e^{-4kt}-2\big)-|\beta|^2,
\end{equation}
for $kt\to\infty,~ X>0$ and the continuously varying function $X$ can change sign and if it passes through zero, 
$X=0$ gives two solutions
\begin{equation}
\label{eq19}
 t_{01,02}=\frac{1}{4k}\log\big[1+2|\beta|^2\mp2\sqrt{|\beta|^4+|\beta|^2-|\alpha|^2}\big].
\end{equation}
For $t_{01,02}$ to be real, we require the condition $|\beta|^4+|\beta|^2-|\alpha|^2>0$. When this condition holds a 
real and positive $t_{02}$ is obtained. $t_{01}$ is also real as
$ 1+2|\beta|^2>2\sqrt{|\beta|^4+|\beta|^2-|\alpha|^2}$, {\it i.e.}, $|\alpha|^2>|\beta|^2$ but $t_{01}$ is positive only for 
$2|\beta|^2>2\sqrt{|\beta|^4+|\beta|^2-|\alpha|^2}$. Thus, we have following two cases:

\begin{itemize}
\item [Case 1]: For $|\alpha|^2>|\beta|^2$, $ 0<t_{01}<t_{02}$ and $X$ takes negative values for time interval $t_{01}<t<t_{02}$.

\item [Case 2]: For $|\alpha|^2\eqslantless|\beta|^2$, $t_{01}\eqslantless0<t_{02}$ and hence $X$ is negative for the interval
$0<t<t_{02}$.
\end{itemize}
To find more confined region of polarization squeezing on time axis we need to look for the time interval for which
$\mathcal{S}_{2}<1$ or $Y<|X|=-X$ as clear from Eq.~(\ref{eq17}). The boundaries of this time interval 
can be given by the roots of equation $Y-|X|=Y+X=0$ as at these times $\mathcal{S}_{2}=1$ and this equation gives the real roots
\begin{equation}
\label{eq20}
t_{1,2}=\frac{1}{4k}\log\big[1+2|\beta|^2\mp 2\sqrt{|\beta|^4-4|\alpha|^2}\big],
\end{equation}
for $|\beta|^4>4|\alpha|^2$. It shows that $0<t_1<t_2$.

It can also be seen that $t_{01}<t_1$ as $ e^{-4kt_1}-e^{-4kt_{01}}=-|\beta|^2-\sqrt{|\beta|^4-4|\alpha|^2}
+2\sqrt{|\beta|^4+|\beta|^2-|\alpha|^2}$ is positive for $|\alpha|^2=0$ and its derivative with 
$|\alpha|^2$, {\it viz.}, $\frac{2}{|\beta|^4-4|\alpha|^2}-\frac{1}{\sqrt{|\beta|^4}+|\beta|^2-|\alpha|^2}>0 $.
We can similarly show that $t_2<t_{02}$ and thus we have $t_01<t_1<t_2<t_{02}$.

Therefore, for $ |\alpha|^2>|\beta|^2$, $ t_{01}>0$ in the interval $ t_{01}<t_1<t_{02}$ in which polarization
squeezing occurs in the interval $t_1<t<t_2$. For $|\alpha|^2\eqslantless|\beta|^2$, however, $ t_{01}\eqslantless0$ 
and hence $X<0$ in the interval $0<t<t_{02}$ in which polarization squeezing occurs in the interval $t_1<t<t_2$.
This becomes clear if we write
\begin{equation}
\label{eq21}
 Y-|X|=\frac{1}{2}e^{-4kt}(e^{4kt}-e^{-4kt_1})(e^{4kt}-e^{-4kt_2}).
\end{equation}
If $x=e^{4kt}-1$ and $x_{1,2}$ are the values of $x$ for $t_1$ and $t_2$, respectively, the expression 
for polarization squeezing factor can be written as 
\begin{equation}
\label{eq22}
 \mathcal{S}_{2}=\frac{\big(4|\alpha|^2+|\beta|^2\big)+x^2}{(x-x_1)(x-x_2)}.
\end{equation}
Minimization of this expression with respect to $x$ by derivative method for a positive root of equation 
$\frac{d(\mathcal{S}_2)}{dx}=0$ gives the minimum value of polarization squeezing factor corresponding to maximum polarization 
squeezing as 
\begin{equation}
\label{eq23}
\mathcal{S}_{2 min}=\frac{\sqrt{1+|\alpha|^2+|\beta|^2}-1}{1+|\beta|^2-\sqrt{1+|\alpha|^2+|\beta|^2}},
\end{equation}
at time
\begin{equation}
\label{eq24}
 t=t_{2min}=\frac{1}{4k}\log \big[{2\sqrt{1+|\alpha|^2+|\beta|^2}-1}\big].
\end{equation}
$\mathcal{S}_{2}$ is less than 1 for $|\beta|^4>4|\alpha|^2$. We thus note that for polarization squeezing to occur 
one shall have two conditions $|\beta|^4>4|\alpha|^2$ and $|\beta|^4>|\alpha|^2-|\beta|^2$. The condition $|\beta|^4>4|\alpha|^2$
is a stronger condition and the other one is certainly followed if this is followed.

\section*{Special case of Equally intense modes}
As a special case of the above, when $|\alpha|^2=|\beta|^2$ , {\it i.e.}, equal number of photons in both the 
polarization modes, we have Eq.~(\ref{eq19}) and Eq.~(\ref{eq20}) reduced in the form
\begin{equation}
\label{eq25}
 t_{01,02}=0, \frac{1}{4k}\log\big[1+|\alpha|^2\big],
\end{equation}
and
\begin{equation}
\label{eq26}
 t_{1,2}=0, \frac{1}{4k}\log\big[1+|\alpha|^2\mp|\alpha|\sqrt{|\alpha|^2-4}\big].
\end{equation}
Polarization squeezing in this case is quantified by polarization squeezing factor
\begin{equation}
\label{eq27}
\mathcal{S}_{2min}=\frac{\sqrt{1+2|\alpha|^2}-1}{\big(1+|\alpha|^2\big)-\sqrt{1+2|\alpha|^2}},
\end{equation}
at the time
\begin{equation}
\label{eq28}
 t=t_{2min}=\frac{1}{4k}\log\big[\big(2\sqrt{1+2|\alpha|^2\big)}-1\big].
\end{equation}
The general condition of polarization squeezing to occur, {\it i.e.} $|\beta|^4>4|\alpha|^2$ reduces to the 
$|\alpha|^2>4$ for this case. \\

\section*{Discussion of Result}
For given number of incident photons, we can have the maximum polarization squeezing
\begin{equation}
\label{eq29}
\mathcal{S}_{2min}=\frac{\sqrt{1+N^2}-1}{1+N^2\sin^2\chi-\sqrt{1+N^2}},\nonumber
\end{equation}
where
$ N=|\alpha|^2+|\beta|^2$ and $\chi= tan^{-1}\big(\frac{|\beta|}{|\alpha|}\big)$.
This gives the further minimized value of polarization squeezing factor exhibiting the maximum polarization squeezing as
\begin{equation}
\label{eq30}
\mathcal{S}_{min}=\frac{1}{\sqrt{1+N^2}},\nonumber
\end{equation}
at time $t_{min}=\frac{1}{4k}\log[2\sqrt{1+N}-1]$ and $\chi=\frac{\pi}{2}$ , {\it i.e.}, $ N=|\beta|^2$ and 
$|\alpha|^2=0$.
It is observed that in order to achieve maximum polarization squeezing the intensity of light should be 
tending to zero in the parametrically amplified mode and maximum in the counter-mode.

\begin{figure}[th]
\centerline{\psfig{file=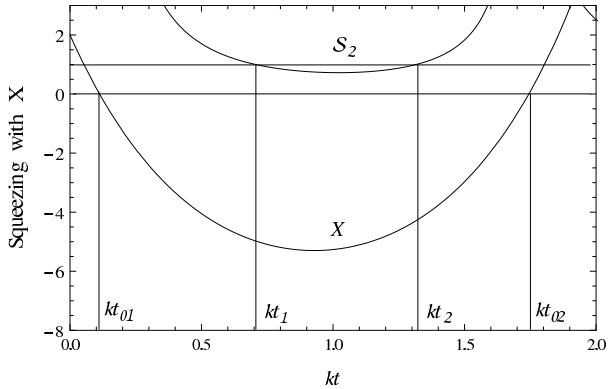,width=8cm}}
\vspace*{8pt}
\caption{Variation of polarization squeezing factor $\mathcal{S}_2$ compared to negativity of $X$ with 
interaction time $kt$, for $|\alpha|^2=10,~|\beta|^2=8$.\label{f1}}
\end{figure}

We found the maximum polarization squeezing is feasible to be analytically calculated in 
the Stokes parameter $\hat S_2 $ by minimization of polarization squeezing factor $\mathcal{S}_2$. It is shown that degree of polarization squeezing is 
maximum for some definite combinations of $\phi_x$ and $\phi_y$ , {\it e.g.}, 
$(\phi_x,\phi_y)=\big(\frac{\pi}{4},\frac{\pi}{4}\big)$ and other values. However, it is not possible to 
show analytically that these are the only maxima. In this particular case, we see analytically that polarization 
squeezing may occur only if the denominator $X$ in the expression for polarization squeezing factor in Eq.~(\ref{eq17}) holds a negative value. 
The variation of polarization squeezing with the negativity of the denominator $X$ is depicted 
in Fig.~\ref{f1} which indicates the time range for occurence of 
polarization squeezing in case of $|\alpha|^2>|\beta|^2$ as an example.  

For polarization squeezing to occur the condition on photon number $|\beta|^4>4|\alpha|^2$ automatically covers the 
condition $|\beta|^4>|\alpha|^2-|\beta|^2$ as $4|\alpha|^2>|\alpha|^2-|\beta|^2$ and it reduces to
$ |\alpha|^2>4$ for equal number of photons in both the polarization modes. It should be noted that this
condition on beam intensity allows to achieve polarization squeezing even for a low beam intensity.

\begin{figure}[th]
\centerline{\psfig{file=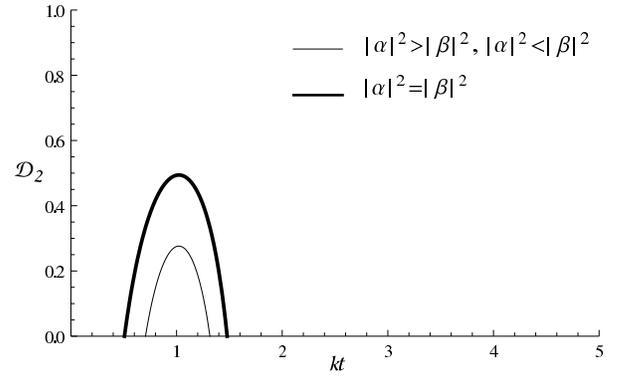,width=8cm}}
\vspace*{8pt}
\caption{Variation of degree of polarization squeezing $\mathcal{D}_2$ with interaction time $kt$.\label{f2}}
\end{figure}

Fig.~\ref{f2} shows the degree of polarization squeezing plotted 
with the interaction time $kt$ as in Eq.~(\ref{eq16}), in all the three possible cases for photon intensities in two polarization modes. 
Some examples of photon number values are studied and the plot shows that the maximum polarization squeezing is 
achieved for equal intensity of light in both the modes. This plot also presents the pattern for occurrence of polarization
squeezing on time axis.

\begin{figure}[th]
\centerline{\psfig{file=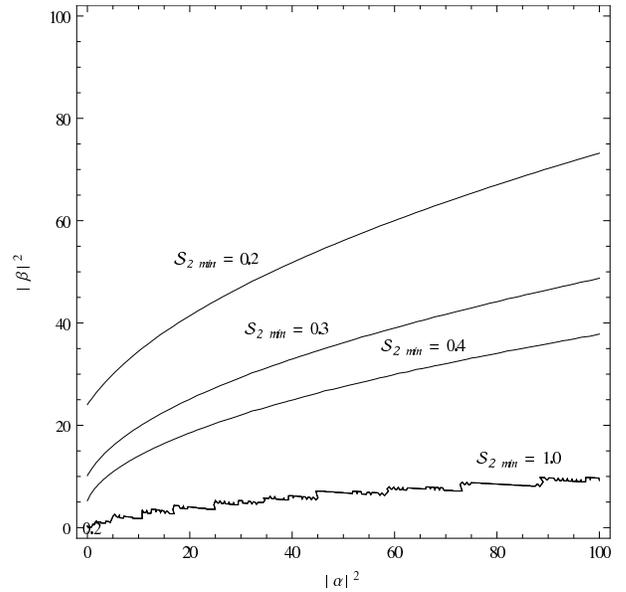,width=8cm}}
\vspace*{8pt}
\caption{Contour plots for differenr values of minimum polarization squeezing factor $\mathcal{S}_{2min}$ 
in $(|\alpha|^2$,~$|\beta|^2)$ plane. $\mathcal{S}_{2min}$ increases with increasing value of 
$|\alpha|^2$ but decreases with $|\beta|^2$.\label{f3}}
\end{figure}

The contour plot for $\mathcal{S}_{2min}$ from Eq.~(\ref{eq16}) in the plane $ (|\alpha|^2, |\beta|^2)$ in 
Fig.~\ref{f3} clearly shows that $\mathcal{S}_{2min}>1$ for $|\beta|^2>2|\alpha|$ which is 
consistent with the essential condition $|\beta|^4>4|\alpha|^2$ for polarization squeezing. 

Efficient polarization squeezing shown to be a resource for quantum communication, achieving good extent of 
polarization squeezing in coherent light via a nonlinear interaction is important. This work gives an added 
advantage of low beam intensity in order to observe polarization squeezing in such a system for significant interaction times under some specific 
conditions. 

\section*{Acknowledgement}
We would like to thank Hari Prakash and Naresh Chandra for their interest and critical comments.


\begin{thebibliography}{0}

\bibitem{1}
G. G. Stokes, {\it Trans.Cambridge Phylos. Soc.} {\bf 9}, (1852) 399..

\bibitem{2} 
M. Born, E. Wolf, {\it Principles of Optics} (Cambridge University Press, England, 1999).

\bibitem{3}
L. Mandel, E. Wolf, {\it Optical Coherence and Quantum Optics Optics} (Cambridge Univ. Press, 
England, 1995).

\bibitem{4} 
H. Prakash, N. Chandra,{\it Physical Review A} {\bf 4}, 796 (1971); 
H. Prakash,and N. Chandra, {\it Physical Review A} {\bf 9}, 1021 (1974); 
H. Prakash,and N. Chandra, {\it Physics Letters A} {\bf 34}, 28 (1971).

\bibitem{5}
U. L. Anderson,and P. Buchhave, {\it Journal of Optics B} {\bf 5}, 486 (2003);
O. Glockl, J. Heersink, N. Korolkova, G. Leuchs, S. Lorenz, {\it Journal of Optics B} {\bf 5}, 492 (2003);
N. Korolkova, R. Loudon, {\it Physical Review A} {\bf 71} 032343 (2005);
N. Korolkova, G. Leuchs, R. Loudon, T. C. Ralph,and Ch. Silberhorn, {Physical Review A} {\bf 65}, 052306 (2002); 
Yu. M. Golubeva, T. Yu. Golubev, M. I. Kolobov, and E. Giacobino, {\it Physical Review A} {\bf 70}, 053817 (2004);
W. P. Bowen, R. Schanabel, H. A. Bachor and P. K. Lam, {\it Physical Review Letters} {\bf 88}, 093601 (2002); 
A. P. Aldojants, S. M. Arakelian, {\it Journal of Modern Optics} {\bf 46}, 475 (1999); 
N. V. Korolkova, A. S. Chirkin, {\it Journal of Modern Optics} {\bf 43}, 869 (1996); 
M. Lassen, M. Sabunch, P. Buchhave, U. L. Anderson, {\it Optics Express} {\bf 15}, 5077 (2007); 
D. N. Klyscho, {\it JETP }{\bf 84}, 1065 (1997);
A. P. Aldojants, S. M. Arakelian, A. S. Chirkin, {\it Quantum Semiclassical Optics} {\bf 9}, 311 (1997);
J. F. Sherson, K. M\o lmer {\it Physical Review Letters} {\bf 97}, 143602 (2006);
J. Heersink, V. Josse, G. Leuchs, V. L. Anderson, {\it Optics Letters} {\bf 30}, 1192 (2005);
P.~Usachev, J.Soderholm, G.~Bjork, A.~Trifonov, Optics Communication, {\bf 193} 161 (2001);
A. P. Aldojants, A. M. Arakelian, A. S. Chirkin, {JETP} {\bf 108}, 63 (1995);
A. S. Chirkin, V. V. Volokhovsky, {\it Journal of Russian Laser Research} {\bf 16}, 6 (1995).


\bibitem{6}
See Ref.~[4].

\bibitem{7}
A. S. Chirkin, A. A. Orlov, and D. Y.Parashchuk, {\it Quantum Electrononics} {\bf 23}, 870 (1993).

\bibitem{8}
J. Heersink, T. Lorenz, O. Glockl, N. Korolkova,and G. Leuchs, {\it Physical Review A} {\bf 68}, 013815 (2003).

\bibitem{9}
A. Luis, N. Korolkova, {\it Physical Review A}, {\bf 74}, 043817 (2006).

\bibitem{10}
R. Prakash and N. Shukla, {\it Optics Communications} {\bf284}, 3568 (2011).

\bibitem{11}
M. O. Schully, M. S. Zubairy,{\it Quantum Optics} (Cambridge Univ. Press, Cambridge, 1997).

\end{thebibliography}
\end{document}